\documentclass[prd,twocolumn,nopacs,floatfix,amsmath,nofootinbib,amssymb,floatfix]{revtex4}
\usepackage{graphicx,color,dcolumn,booktabs,bm}
\usepackage{longtable,lscape}
\usepackage{txfonts}
\usepackage{overpic}
\usepackage{amssymb}
\usepackage{indentfirst}
\usepackage{feynmf}   
\usepackage{slashed}  
\usepackage{cases}
\usepackage{color}
\usepackage{multirow}
\usepackage{epstopdf}
\usepackage{longtable}
\usepackage{ulem}
\usepackage{graphicx,color,dcolumn,booktabs,bm}
\usepackage[colorlinks,
            citecolor=blue,
            anchorcolor=red,
            menucolor=red,
            linkcolor=red,
            filecolor=red,
            runcolor=red,
            urlcolor=blue,
            frenchlinks=red]{hyperref}

\graphicspath{{Figures/}} %

\allowdisplaybreaks

\begin{document}

\title{Exploring the mass spectrum and the electromagnetic properties of the possible $\Xi_{cc}K^{(*)}$ and $\Xi_{cc}\bar{K}^{(*)}$ molecules}

\author{Li-Cheng Sheng$^{1}$}
\author{Jin-Yu Huo$^{1}$}
\author{Rui Chen$^{1}$\footnote{Corresponding author}}\email{chenrui@hunnu.edu.cn}
\author{Fu-Lai Wang$^{2,3,4,6}$}
\author{Xiang Liu$^{2,3,4,5,6}$}
\email{xiangliu@lzu.edu.cn}

\affiliation{
$^1$Key Laboratory of Low-Dimensional Quantum Structures and Quantum Control of Ministry of Education, Department of Physics and Synergetic Innovation Center for Quantum Effects and Applications, Hunan Normal University, Changsha 410081, China\\
$^2$School of Physical Science and Technology, Lanzhou University, Lanzhou 730000, China\\
$^3$Lanzhou Center for Theoretical Physics, Key Laboratory of Theoretical Physics of Gansu Province, Lanzhou University, Lanzhou 730000, China\\
$^4$Key Laboratory of Quantum Theory and Applications of MoE, Lanzhou University, Lanzhou 730000, China\\
$^5$MoE Frontiers Science Center for Rare Isotopes, Lanzhou University, Lanzhou 730000, China\\
$^6$Research Center for Hadron and CSR Physics, Lanzhou University and Institute of Modern Physics of CAS, Lanzhou 730000, China}
\date{\today}

\begin{abstract}
Using the one-boson-exchange model, we investigate the interactions between the doubly charmed baryon $\Xi_{cc}(3621)$ and the $S-$wave (anti-)kaon accounting for the $S-D$ wave mixing and coupled-channel effects. We find the coupled $\Xi_{cc}K/\Xi_{cc}K^*$ state with $I(J^P)=0(1/2^-)$, the $\Xi_{cc}K^*$ state with $0(1/2^-)$, the $\Xi_{cc}\bar{K}$ state with $0(1/2^-)$, and the $\Xi_{cc}\bar{K}^*$ states with $0(1/2^-,3/2^-)$ can be recommended as good doubly charmed molecular candidates with strangeness $|S|=1$. We further examine their M1 radiative decay behaviors and magnetic moments within the constituent quark model framework. This information can enhance our understanding of their inner structures, including the distribution of electric charge and the orientation of the constituent quarks' spins.
\end{abstract}


\maketitle

\section{Introduction}

In recent decades, experimental discoveries of new hadron matter have significantly advanced the study of hadronic spectroscopy, making it a prominent research focus in particle physics (see reviews \cite{Chen:2016qju,Liu:2019zoy,Chen:2016spr,Guo:2017jvc,Chen:2022asf,Liu:2013waa,Hosaka:2016pey} for more details). Delving deeply into hadronic spectroscopy cannot only reveal the inner structures and formation mechanisms of new hadronic states but also provide insights into the complex nonperturbative dynamics of quantum chromodynamics.

Recently, the LHCb Collaboration observed two open-charm exotic structures, $T_{c\bar{s}}^{a0(++)}(2900)$ in the $D_s^+ \pi^-(\pi^+)$ final states \cite{LHCb:2022sfr}. Their quantum number configurations are $I(J^P)=1(0^+)$. The masses and decay widths are $M_{T_{c\bar{s}}^{a0}(2900)}=2892 \pm 14 \pm 15$ MeV, $\Gamma_{T_{c\bar{s}}^{a0}(2900)}=119 \pm 26 \pm 12$ MeV, $M_{T_{c\bar{s}}^{a++}(2900)}=2921 \pm 17 \pm 19$ MeV, and $\Gamma_{T_{c\bar{s}}^{a++}(2900)}=137 \pm 32 \pm 14$ MeV. Due to their near-threshold behaviors and quantum numbers, the $T_{c\bar{s}}^{a0(++)}(2900)$ are proposed as isovector $D^\ast K^\ast$ molecules with $J^P=0^+$. The hadronic molecular explanations remind us of the famous charm-strange states, the $D_{s0}(2317)$ and $D_{s1}(2460)$. These states were observed by the $BaBar$ and CLEO experiments \cite{BaBar:2003oey,CLEO:2003ggt} in the $D_s^{+}\pi^0$ and $D_s^{*+}\pi^0$ decay modes, respectively. These states were further confirmed by the Belle and $BaBar$ experiments \cite{Belle:2003kup,BaBar:2006eep,BaBar:2003cdx}. Theoretically, since they are located below the $DK$ and $D^*K$ mass thresholds, $D_{s0}(2317)$ and $D_{s1}(2460)$ are often interpreted as $S$-wave isoscalar $DK$ and $D^*K$ molecular states, respectively \cite{Barnes:2003dj,Chen:2004dy,Guo:2006fu,Guo:2006rp,Navarra:2015iea,Faessler:2007us,Xie:2010zza,
Feng:2012zze,Yang:2021tvc}.

Additionally, it is worth mentioning that in 2020, the LHCb Collaboration first observed two other structures in the $D^-K^+$ invariant mass distribution, namely, the $X_0(2900)$ and $X_1(2900)$, by analyzing the decay amplitude for the $B^+\to D^+D^-K^+$ process \cite{LHCb:2020pxc}. The corresponding masses and decay widths are $M_{X_0(2900)}=2866\pm7\pm2$ MeV, $\Gamma_{X_0(2900)}=57\pm12\pm4$ MeV, $M_{X_1(2900)}=2904\pm5\pm1$ MeV, and $\Gamma_{X_1(2900)}=110\pm11\pm4$ MeV, respectively. Their spin-parities are $J^P=0^+$ and $1^-$, respectively. The $X_0(2900)$ and $X_1(2900)$ are also good candidates for open-charm tetraquark states, as the valence quark components in the final states are $[\bar{c}suu]$. Because the $X_{0,1}(2900)$ are very close to the mass thresholds for the $\bar{D}^*K^*$ and $\bar{D}_1K$ systems, they are often regarded as isoscalar meson-meson hadron molecules \cite{Molina:2020hde,Kong:2021ohg,Wang:2021lwy,Chen:2020aos,Agaev:2020nrc,He:2020btl,Qi:2021iyv,Chen:2021tad}.

In the bottom sector, there is an interesting discussion on the existence of open-bottom tetraquarks. Experimentally, in 2016, the D\O\, Collaboration reported a narrow structure, $X(5568)$, in the $B_s^0\pi^\pm$ invariant mass spectrum \cite{D0:2016mwd}. However, it was not confirmed by the ATLAS, CDF, CMS, and LHCb Collaborations \cite{ATLAS:2018udc,CDF:2017dwr,CMS:2017hfy,LHCb:2016dxl}. Theoretically, the hadronic explanations for the $X(5568)$ have been examined in several studies \cite{Xiao:2016mho,Agaev:2016urs,Burns:2016gvy,Albaladejo:2016eps,Chen:2016ypj,Lu:2016kxm,Sun:2016tmz}. In particular, in our previous work \cite{Chen:2016ypj}, we found that the $X(5568)$ cannot be assigned as an isovector $B\bar{K}$ molecular state, however, there can exist possible isoscalar $B^{(*)}\bar{K}^{(*)}$ molecular states. In addition, the $B_{sJ}(6158)$ was explained as a $\bar{B}K^*$ molecular state with $1^+$ \cite{Kong:2021ohg}, and the masses for the predicted molecules are also in consonance with the results in Refs.\cite{Zanetti:2016wjn,Dias:2016dme}.

\begin{eqnarray*}
\renewcommand\tabcolsep{0.25cm}
\renewcommand{\arraystretch}{1.32}
\begin{array}{ccc}
\toprule[1pt]
\text{Single charm}\rightleftharpoons     &\text{Double charm}\rightleftharpoons    &\text{Single bottom}\\
\hline\nonumber
D_{s0}(2317) \sim DK      &P_{cc\bar{s}} \sim \Xi_{cc}K^{(*)}    &X(5568)\sim \bar B{K}\\
D_{s1}(2460) \sim D^*K        &&B_{sJ}(6158)\sim \bar BK^*\\
T_{c\bar s}(2900) \sim D^*K^*     &   &\\
X(2900) \sim \bar{D}^*K^*     &P_{cc{s}} \sim \Xi_{cc}\bar K^{(*)}   &\\
\bottomrule[1pt]\bottomrule[1pt]
\end{array}
\end{eqnarray*}

As shown above, it is very interesting to explore the $\Xi_{cc}K^{(*)}$ and $\Xi_{cc}\bar{K}^{(*)}$ interactions to search for possible doubly charmed molecular candidates with strangeness $|S|=1$. In fact, if we ignore the coupling between the heavy quarks and the light quarks, the interactions between the double charmed baryon and kaon are very similar to the interactions between the charmed meson and kaon. Due to the heavier reduced masses, there is a potential for the existence of the $\Xi_{cc}K^{(*)}$ and $\Xi_{cc}\bar{K}^{(*)}$ bound states.

In this work, we adopt the one-boson-exchange (OBE) model to derive the effective potentials for the $\Xi_{cc}K^{(*)}$ and $\Xi_{cc}\bar{K}^{(*)}$ systems, including the $\pi$, $\sigma$, $\eta$, $\rho$, and $\omega$ exchange interactions. Here, we also consider the $S-D$ wave mixing effects and the coupled-channel effects. Indeed, other channels, such as $\Lambda_{c}D_s^{(*)}$ and $\Sigma_{c}D_s^{(*)}$, involving the charmed baryon and charm-strange meson, can couple to the $\Xi_{cc}K^{(*)}$ systems. This is because their mass thresholds lie in the same energy region as the $\Xi_{cc}K^{(*)}$ systems, and they share the same quantum numbers. However, the interactions from $\Xi_{cc}K^{(*)}\to \Lambda_{c}D_s^{(*)}$ and $\Xi_{cc}K^{(*)}\to \Sigma_{c}D_s^{(*)}$ processes, which arise from exchanging charmed mesons are very short-ranged. These interactions play a minor role in generating loosely bound moecular states. Therefore, we will omit the consideration of these coupled interactions in the subsequent discussion.

Using these OBE effective potentials, we solve the coupled channel Schr\"{o}dinger equations to search for possible loosely bound state solutions. Our study can not only predict the existence of possible $\Xi_{cc}K^{(*)}$ and $\Xi_{cc}\bar{K}^{(*)}$ molecular candidates, but also deepen our understanding of the interactions between the doubly charmed baryon and the (anti-)strange meson. In addition, this study will test the molecular-state picture for the open-charm exotic structures with strangeness to some extent.

Afterward, we use the obtained bound state solutions to explore the M1 radiative decay behaviors and the magnetic moments of the predicted molecular candidates. And we employ the constituent quark model, which is often adopted to study the electromagnetic properties of the baryons \cite{Schlumpf:1993rm,Cheng:1997kr,Majethiya:2009vx,Sharma:2010vv,Ghalenovi:2014swa,
Girdhar:2015gsa,Thakkar:2016sog,Dahiya:2018ahb,Mohan:2022sxm} and multiquarks \cite{Liu:2003ab,Huang:2004tn,Zhu:2004xa,Wang:2016dzu,Li:2021ryu,Wang:2022tib,Zhou:2022gra,
Gao:2021hmv,Wang:2022nqs,Wang:2023aob,An:2022qpt,Wu:2022gie,Guo:2023fih,Wang:2023bek,Wang:2023ael,
Li:2024wxr,Li:2024jlq,Lai:2024jfe}. These efforts can provide valuable clues to understand the inner structures of the open heavy flavor multiquarks.

This paper is organized as follows. In Sec.~\ref{sec2}, we introduce the theoretical frameworks of the OBE effective potentials, the M1 radiative decay behaviors, and the magnetic moments for the $\Xi_{cc}K^{(*)}$ and $\Xi_{cc}\bar{K}^{(*)}$ systems. In Sec.~\ref{sec3}, we present the corresponding numerical results. The paper ends with a summary in Sec. \ref{sec4}.

\section{Theoretical frameworks}\label{sec2}

\subsection{OBE effective potentials}

In this subsection, we study the $\Xi_{cc}K^{(*)}$ and $\Xi_{cc}\bar{K}^{(*)}$ interactions using the OBE model. First, we introduce the relevant Lagrangians. For the interactions between the strange mesons and the light mesons $(\sigma, \pi, \eta, \rho, \omega)$, we construct the effective Lagrangians based on $SU(3)$ symmetry \cite{Lin:1999ad,Nagahiro:2008mn,Chen:2019uvv}, i.e.,
\begin{eqnarray}
\mathcal{L}_{PP\sigma} &=& g_\sigma m_P \langle PP\sigma\rangle,\label{lag01}\\
\mathcal{L}_{VV\sigma} &=& g_\sigma m_V \langle VV\sigma\rangle,\\
\mathcal{L}_{PPV} &=& \frac{ig}{\sqrt{2}}\langle\partial^{\mu}{P}(PV_{\mu}-V_{\mu}P)\rangle,\\
\mathcal{L}_{VVP} &=&  \frac{g_{VVP}}{\sqrt{2}}\varepsilon^{\mu\nu\alpha\beta}\langle\partial_{\mu}{V_{\nu}}\partial_{\alpha}{V_{\beta}}P\rangle,\label{lag02}\\
\mathcal{L}_{VVV} &=& \frac{ig}{2\sqrt{2}}\langle\partial^{\mu}{V^{\nu}}(V_{\mu}V_{\nu}-V_{\nu}V_{\mu})\rangle.\label{lag03}
\end{eqnarray}
Here, $P$ and $V$ are the $SU(3)$ matrices containing
the octets of vector and pseudoscalar mesons, respectively, which are given as follows:
\begin{eqnarray*}
P &=& \left(\begin{array}{ccc}
\frac{\pi^0}{\sqrt{2}}+\frac{\eta}{\sqrt{6}} &\pi^+ &K^+ \nonumber\\
\pi^- &-\frac{\pi^0}{\sqrt{2}}+\frac{\eta}{\sqrt{6}} &K^0 \nonumber\\
K^- &\bar{K}^0 &-\sqrt{\frac{2}{3}}\eta
\end{array}\right),\nonumber\\
V &=& \left(\begin{array}{ccc}
\frac{\rho^0}{\sqrt{2}}+\frac{\omega}{\sqrt{2}} &\rho^+ &K^{*+} \nonumber\\
\rho^- &-\frac{\rho^0}{\sqrt{2}}+\frac{\omega}{\sqrt{2}} &K^{*0} \nonumber\\
K^{*-} &\bar{K}^{*0} &\phi
\end{array}\right).
\end{eqnarray*}

Once expanding the Eqs. (\ref{lag01})$-$(\ref{lag03}), we can obtain
\begin{eqnarray}
\mathcal{L}_{K^{(*)}K^{(*)}\sigma} &=& g^{\prime}_{\sigma}m_K\bar{K}K\sigma-g^{\prime}_{\sigma}m_{K^*}\bar{K^*}\cdot{K^*}\sigma,\label{lag04}\\
\mathcal{L}_{PKK^*} &=&
        \frac{ig}{4}\left[\left(\bar{K}^{*\mu}K-\bar{K}K^{*\mu}\right)\left(\bm{\tau}\cdot{\partial_\mu{\bm{\pi}}}+\frac{\partial_\mu{\eta}}{\sqrt{3}}\right)\right]\nonumber\\
        &&+\frac{ig}{4}\left[\left(\partial_\mu{\bar{K}}K^{*\mu}-\bar{K}^{*\mu}\partial_\mu{K}\right)\left(\bm{\tau}\cdot{\bm{\pi}}+\frac{\eta}{\sqrt{3}}\right)\right],\label{lag05}\\
\mathcal{L}_{VKK} &=&
        \frac{ig}{4}\left[\bar{K}\partial_\mu{K}-\partial_\mu{\bar{K}K}\right]\left(\bm{\tau}\cdot{\bm{\rho}^\mu}+\omega^\mu\right),\label{lag06}\\
\mathcal{L}_{VK^*K^*} &=&
\frac{ig}{4}\left[\left(\bar{K}^*_\mu\partial^\mu{K}^{*\nu}-\partial^\mu{\bar{K}^{*\nu}}K^*_\mu\right)\left(\bm{\tau}\cdot{\bm{\rho}_\nu}+\omega_\nu\right)\right]\nonumber\\
&&+\frac{ig}{4}\left[\left(\partial^\mu{\bar{K}}^{*\nu}K^*_\nu-\bar{K}^*_\nu\partial^\mu{\bar{K}^{*\nu}}\right)\left(\bm{\tau}\cdot{\bm{\rho}_\mu}+\omega_\mu\right)\right]\nonumber\\
&&+\frac{ig}{4}\left[\left(\bar{K}^*_{\nu}K^*_\mu-\bar{K}^*_{\mu}K^*_\nu\right)
\left(\bm{\tau}\cdot{\partial^\mu{\bm{\rho}}^\nu+\partial^\mu{\omega}^\nu}\right)\right],\label{tag07}\\
\mathcal{L}_{PK^*K^*} &=&
g_{VVP}\varepsilon_{\mu\nu\alpha\beta}\partial^\mu\bar{K}^{*\nu}\partial^\alpha{K}^{*\beta}\left(\bm{\tau}\cdot{\bm{\pi}}+\frac{\eta}{\sqrt{3}}\right),\label{tag08}\\
\mathcal{L}_{VKK^*} &=&
g_{VVP}\varepsilon_{\mu\nu\alpha\beta}\left(\partial^\mu{\bar{K}^{*\nu}}K+\bar{K}\partial^\mu{K}^{*\nu}\right)\nonumber\\
&&\left(\bm{\tau}\cdot{\partial^\alpha{\bm{\rho}^\beta}}+\partial^\alpha{\omega^\beta}\right).\label{tag09}
\end{eqnarray}

Drawing from the experience from the nucleon-nucleon interactions, we construct the effective Lagrangians between the doubly charmed baryons and the light mesons as follows \cite{Meng:2017fwb,Meng:2017udf}:
\begin{eqnarray}
\mathcal{L}_{\Xi_{cc}\Xi_{cc}\sigma}&=&g_{\sigma}^{\prime}\bar{\Xi}_{cc}\sigma\Xi_{cc},\label{lag10}\\
\mathcal{L}_{\Xi_{cc}\Xi_{cc}P}&=&g_{\pi}\bar{\Xi}_{cc}i\gamma_5P\Xi_{cc},\label{lag11}\\
\mathcal{L}_{\Xi_{cc}\Xi_{cc}V}&=&h_{v}\bar{\Xi}_{cc}\gamma_{\mu}V^{\mu}\Xi_{cc}
+\frac{f_v}{2m_{\Xi_{cc}}}\bar{\Xi}_{cc}\sigma_{\mu\nu}\partial^{\mu}{V^{\nu}}\Xi_{cc}.\label{lag14}
\end{eqnarray}

We estimate the coupling constants in Eqs. (\ref{lag01})$-$(\ref{lag14}) as follows. With the help of the the quark model, we can obtain the relationship between the coupling constants of double-charm baryons and those of  nucleon-nucleon interaction \cite{Machleidt:2000ge,Machleidt:1987hj,Cao:2010km},
$g_{\sigma} = \frac{1}{3}g_{{\sigma}NN}$, $g_\pi = -\frac{\sqrt{2}m_{\Xi_{cc}}}{5m_N}g_{{\pi}NN}$, $h_v = \sqrt{2}g_{{\rho}NN}$, and $h_v+f_v = -\frac{\sqrt{2}m_{\Xi_{cc}}}{5m_N}(g_{{\rho}NN}+f_{{\rho}NN})$.Note that $g^{\prime}_\sigma = -3.65$ \cite{Chen:2023qlx}, $g = 12.0$ \cite{Chen:2017xat}, $g_{VVP} = 3g^2/(32\sqrt{2}\pi^2f_{\pi})$, {and} $f_\pi = 132$ MeV are estimated by using the vector meson dominance model \cite{Kaymakcalan:1983qq}. We further fix the phase between all the coupling constants in the quark model. These values are summarized in Table \ref{tab01}.

\begin{table}[!hbtp]
\renewcommand\tabcolsep{0.3cm}
\renewcommand{\arraystretch}{1.4}
\caption{Coupling constants adopted in our calculation. The coupling constants for the nucleon-nucleon interactions are taken from Refs. \cite{Machleidt:2000ge,Machleidt:1987hj,Cao:2010km}.}\label{tab01}
\begin{tabular}{cccc}
\toprule[1pt]\toprule[1pt]
$\sigma$    &$\pi/\eta$  &$\rho/\omega$ & \\ \midrule[1pt]
$\frac{g^2_{{\sigma}NN}}{4\pi}=5.69$ &$\frac{g^2_{{\pi}NN}}{4\pi}=13.60$ &$\frac{g^2_{{\rho}NN}}{4\pi}=0.84$&$\frac{f_{{\rho}NN}}{g_{{\rho}NN}}=6.10$\\
$g_{\sigma}=-2.82$&$g_{\pi}=14.26$&$h_v=4.65$&$f_v=-28.11$\\\bottomrule[1pt]\bottomrule[1pt]
\end{tabular}
\end{table}

With the preparations of the effective Lagrangians, we next derive the OBE effective potentials for the $\Xi_{cc}K^{(*)}$ systems, which relate to the scattering amplitudes in the Breit approximation:
\begin{eqnarray}\label{breit}
\mathcal{V}_{E}^{h_1h_2\to h_3h_4}(\bm{q}) &=&
          -\frac{\mathcal{M}(h_1h_2\to h_3h_4)}
          {\sqrt{\prod_i2M_i\prod_f2M_f}}.
\end{eqnarray}
Here, $\mathcal{M}(h_1h_2\to h_3h_4)$ denotes the scattering amplitude for the $h_1h_2\to h_3h_4$ process by exchanging the light meson $(\sigma, \pi, \eta, \rho, \omega)$ in $t$ channel. $M_i$ and $M_f$ are the masses of the initial states ($h_1$, $h_2$) and final states ($h_3$, $h_4$), respectively. By performing a Fourier transformation, we can obtain the effective potential in the coordinate space $\mathcal{V}(\bm{r})$:
\begin{eqnarray}
\mathcal{V}_{E}^{h_1h_2\to h_3h_4}(\bm{r}) =
          \int\frac{d^3\bm{q}}{(2\pi)^3}e^{i\bm{q}\cdot\bm{r}}
          \mathcal{V}_{E}^{h_1h_2\to h_3h_4}(\bm{q})\mathcal{F}^2(q^2,m_E^2).\nonumber
\end{eqnarray}
Here, we introduce a monopole form factor $\mathcal{F}(q^2,m_E^2) = (\Lambda^2-m_E^2)/(\Lambda^2-q^2)$ at each interactive vertex, which can reflect the off-shell effect of the exchanged boson. $\Lambda$, $m_E$, and $q$ are the cutoff, mass, and four-momentum of the exchanged meson, respectively. In our OBE effective potentials, the cutoff $\Lambda$ is the only free parameter, relating to the typical hadronic scale or the intrinsic size of hadrons. Based on the experience of the deuteron \cite{Tornqvist:1993ng,Tornqvist:1993vu}, a reasonable cutoff value is around $\Lambda \sim 1.00$ GeV.

In order to obtain the total OBE effective potentials, we further construct the wave functions for the investigated systems. For the $\Xi_{cc}K^{(*)}$ systems, the wave function $|\Psi\rangle$ can be constructed as the direct product of the color wave function $|\phi_c\rangle$, the flavor wave function $|I,I_3\rangle$, the spin-orbit wave function $|{}^{2S+1}L_J\rangle$, and the radial wave function $|\psi(r)\rangle$:
\begin{eqnarray}
|\Psi\rangle &=& |\phi_c\rangle|I,I_3\rangle|{}^{2S+1}L_J\rangle|\psi(r)\rangle. \label{wave}
 \end{eqnarray}
Here, the color wave function is 1, indicating that the color configuration is a singlet. The isospin $I$ can be either 0 or 1 for the $\Xi_{cc}K^{(*)}$ systems. We can construct their flavor function as follows:
\begin{eqnarray}
|0,0\rangle&=&\sqrt{\frac{1}{2}}\left(\left|\Xi_{cc}^{++}K^{(*)0}\right\rangle-\left|\Xi_{cc}^{+}K^{(*)+}\right\rangle\right),\nonumber\\
|1,1\rangle&=&\left|\Xi_{cc}^{++}K^{(*)+}\right\rangle,\nonumber\\
|1,0\rangle&=&\sqrt{\frac{1}{2}}\left(\left|\Xi_{cc}^{++}K^{(*)0}\right\rangle+\left|\Xi_{cc}^{+}K^{(*)+}\right\rangle\right),\nonumber\\
|1,-1\rangle&=&\left|\Xi_{cc}^{+}K^{(*)0}\right\rangle.\nonumber
\end{eqnarray}

When we consider the $S-D$ wave mixing effects, the spin-orbit wave functions $|{}^{2S+1}L_J\rangle$ can include
\begin{eqnarray}
\Xi_{cc}K:    &J^P=\frac{1}{2}^-&  \left|{}^{2}S_{\frac{1}{2}}\right\rangle,\label{lag15}\\
\Xi_{cc}K^*:  &J^P=\frac{1}{2}^-&  \left|{}^{{2}}S_{\frac{1}{2}}\right\rangle,\quad \left|{}^4D_{\frac{1}{2}}\right\rangle,\label{lag16}\\
&J^P=\frac{3}{2}^-&  \left|{}^4S_\frac{3}{2}\right\rangle,\quad   \left|{}^2D_\frac{3}{2}\right\rangle,\quad
                                      \left|{}^4D_\frac{3}{2}\right\rangle.\label{lag17}
\end{eqnarray}
The general expressions can be expressed as follows:
\begin{eqnarray}
\Xi_{cc}K:\left|{}^{2S+1}L_J\right\rangle&=&
\chi_{\frac{1}{2}m_s}\left|Y_{L,m_L}\right\rangle,\nonumber\\
\Xi_{cc}K^*:\left|{}^{2S+1}L_J\right\rangle&=&
\sum_{m,m^\prime}^{m_s,m_L}C^{S,m_S}_{\frac{1}{2}m,1m^\prime}
C^{J,M}_{Sm_S,Lm_L}\chi_{\frac{1}{2}m}\epsilon^{m^\prime}\left|Y_{L,m_L}\right\rangle.\nonumber
\end{eqnarray}
Here, $C^{S,m_S}_{\frac{1}{2}m,1m^\prime}$ and $C^{J,M}_{Sm_S,Lm_L}$ are the Clebsch-Gordan coefficients. $\chi_{\frac{1}{2}m}$ and $\left|Y_{L,m_L}\right\rangle$ denote the spin wave function and spherical harmonics function, respectively. $\epsilon^{m^\prime}$ is the polarization vector for the vector meson $K^*$, which has the form of $\epsilon_{\pm}^{m}=\mp\frac{1}{\sqrt{2}}\left(\epsilon_x^{m}{\pm}i\epsilon_y^{m}\right)$ and $\epsilon_0^{m}=\epsilon_z^{m}$, with $\epsilon_{\pm1}= \frac{1}{\sqrt{2}}\left(0,\pm1,i,0\right)$ and $\epsilon_{0} =\left(0,0,0,-1\right)$.

Finally, the total OBE effective potentials for the $\Xi_{cc}K\to\Xi_{cc}K$, $\Xi_{cc}K^*\to\Xi_{cc}K^*$, and $\Xi_{cc}K\to\Xi_{cc}K^*$ processes, denoted as $V_{11}$, $V_{22}${\color{blue}\uwave{,}} and $V_{12}$, respectively, can be written as
\begin{widetext}
\begin{eqnarray}
V_{11} &=& -\frac{1}{2}g_{\sigma}g_{\sigma}^{\prime}Y(\Lambda,m_{\sigma},r)
     +\frac{\sqrt{2}}{8}gh_V\mathcal{Y}^I_1(\Lambda,m_\rho,m_\omega,r)
     +\frac{\sqrt{2}f_{V}g}{32m^{2}_{\Xi_{cc}}}
     \nabla^2\mathcal{Y}^I_1(\Lambda,m_\rho,m_\omega,r),\\\label{potential1}
V_{22} &=& -\frac{1}{2}g_\sigma g_{\sigma}^{\prime}(\epsilon^{\dag}_4\cdot\epsilon_2) Y(\Lambda,m_{\sigma},r)
     +\frac{\sqrt{2} g_\pi g_{VVP}}{24m_{\Xi_{cc}}}\mathcal{F}(r,i\bm{\sigma},\bm{\epsilon}_2\times\bm{\epsilon}_4^\dag)
     \mathcal{Y}^I_2(\Lambda,m_\pi,m_\eta,r)
     +\frac{\sqrt{2}h_V g}{8}(\bm{\epsilon_{4}^\dag}\cdot\bm{\epsilon_{2}})
     \mathcal{Y}^I_1(\Lambda,m_\rho,m_\omega,r)\nonumber\\
     &&+\frac{\sqrt{2}h_V g}{32 m^{2}_{\Xi_{cc}}}(\bm{\epsilon_{4}^\dag}\cdot\bm{\epsilon_{2}})
     \nabla^2\mathcal{Y}^I_1(\Lambda,m_\rho,m_\omega,r)
     +\frac{\sqrt{2}}{64}\frac{(h_V+f_V)g}{m_{\Xi_{cc}}m_{K^{*}}}
     \mathcal{F}^\prime(r,i\bm{\sigma},\bm{\epsilon}_2\times\bm{\epsilon}_4^\dag)
     \mathcal{Y}^I_1(\Lambda,m_\rho,m_\omega,r),\nonumber\\\\\label{lag20}
V_{12}&=&   -\frac{\sqrt{2}}{24}\frac{h_Vg_{VVP}}{24m_{\Xi_{cc}}}\sqrt{\frac{m_{K^*}}{m_K}}
\mathcal{F}^\prime(r,\bm{\sigma},\bm{\epsilon}_4^\dag)\mathcal{Y}^I_1(\Lambda_0,m_{\rho0},m_{\omega0},r)-\frac{\sqrt{2}}{24}\frac{f_Vg_{VVP}}{24m_{\Xi_{cc}}}\sqrt{\frac{m_{K^*}}{m_K}}
\mathcal{F}^\prime(r,\bm{\sigma},\bm{\epsilon}_4^\dag)
\mathcal{Y}^I_1(\Lambda_0,m_{\rho0},m_{\omega0},r)\nonumber\\    &&-\frac{\sqrt{2}}{64}\frac{g_{\pi} g}{m_{\Xi_{cc}}\sqrt{m_K m_{K^*}}}\mathcal{F}(r,\bm{\sigma},\bm{\epsilon}_4^\dag)
\mathcal{Y}^I_2(\Lambda_0,m_{\pi0},m_{\eta0},r).\nonumber\\\label{potential3}
\end{eqnarray}
\end{widetext}
Here, the variables in the Eq. (\ref{potential3}) are $\Lambda_0^2=\Lambda^2-q_0^2$, $m_0^2=m^2-q_0^2$, and $q_0=\frac{m_{K^*}^2-m_{K}^2}{2(m_{K^*}+m_{\Xi_{cc}})}$. And we define serval useful functions:
\begin{eqnarray}
\mathcal{F}(r,\bm{a},\bm{b}) &=& \bm{a}\cdot\bm{b}\nabla^2
     +S(\hat{r},\bm{a},\bm{b})
     r\frac{\partial}{\partial r}\frac{1}{r}\frac{\partial}{\partial r},\nonumber\\
\mathcal{F}^{\prime}(r,\bm{a},\bm{b}) &=&
     2\bm{a}\cdot\bm{b}\nabla^2
     -S(\hat{r},\bm{a},\bm{b})
     r\frac{\partial}{\partial r}\frac{1}{r}\frac{\partial}{\partial r},\nonumber\\
S(\hat{r},\bm{a},\bm{b}) &=& 3(\hat{r}\cdot\bm{a})(\hat{r}\cdot\bm{b})-\bm{a}\cdot\bm{b},\nonumber\\
\mathcal{Y}^I_1(\Lambda,m_1,m_2,r) &=& \mathcal{G}(I)Y(\Lambda,m_1,r)+Y(\Lambda,m_1,r),\nonumber\\
\mathcal{Y}^I_2(\Lambda,m_1,m_2,r) &=& \mathcal{G}(I)Y(\Lambda,m_2,r)+\frac{1}{3}Y(\Lambda,m_2,r),\nonumber
\end{eqnarray}
with $Y(\Lambda,m,r) = \frac{e^{-mr}-e^{-\Lambda r}}{4\pi r}-\frac{\Lambda^2-m^2}{8\pi \Lambda}e^{-\Lambda r}$, the isospin factors $\mathcal{G}(0) =-3$ and $\mathcal{G}(1) =1$. The $\bm{a}\cdot\bm{b}$ and $S(\hat{r},\bm{a},\bm{b})$ are the spin-spin interactions and tensor force operators, respectively. In Table \ref{tab02}, we collect the matrices elements for the spin-spin interactions and tensor force operators involved in the OBE effective potentials as shown in Eqs. (\ref{potential1})$-$(\ref{potential3}), they can be obtained by sandwiching the discussed spin-orbit wave functions in Eqs. (\ref{lag15})$-$(\ref{lag17}).

\begin{table}[!htbb]
\renewcommand\tabcolsep{0.3cm}
\renewcommand{\arraystretch}{1.15}
\caption{Matrices elements for the spin-spin interactions and tensor force operators in the OBE effective potentials.}\label{tab02}
{\begin{tabular}{lcc}
\toprule[1pt]\toprule[1pt]
{}
       &$J=\frac{1}{2}$      &$J=\frac{3}{2}$\\\hline
$\langle\bm{\epsilon_4^\dag}\cdot\bm{\epsilon_2}\rangle$   &$\bigl(\begin{smallmatrix}1&0\\0&1\end{smallmatrix}\bigl)$
                  &\multirow{3}{*}{$\left(\begin{array}{ccc}1  &0  &0\\  0   &1   &0\\0  &0   &1\end{array}\right)$} \\\\\\
$\langle i\bm{\sigma}\bm{\cdot}\left(\bm{\epsilon_2}\times\bm{\epsilon_4^{\dag}}\right)\rangle$    &$\bigl(\begin{smallmatrix}-2&0\\0&1\end{smallmatrix}\bigl)$
             &\multirow{3}{*}{$\left(\begin{array}{ccc}1  &0  &0\\  0   &-2   &0\\0  &0   &1\end{array}\right)$} \\\\\\
$\langle S(\hat{r},\bm{\sigma,i\epsilon_2\times\epsilon_4^\dag})\rangle$    &$\bigl(\begin{smallmatrix}0&-\sqrt{2}\\-\sqrt{2}&-\sqrt{2}\end{smallmatrix}\bigl)$
                   &\multirow{3}{*}{$\left(\begin{array}{ccc}0  &1  &2\\  1   &0   &-1\\2  &-1   &0\end{array}\right)$} \\\\\\
$\langle \bm{\sigma}\bm{\cdot}\bm{\epsilon_4^\dag}\rangle$    &$\bigl(\begin{smallmatrix}\sqrt{3}\\0\end{smallmatrix}\bigl)$
                   &\ldots\\
$\langle S(\hat{r},\bm{\sigma},\bm{\epsilon_4^\dag})\rangle$    &$\bigl(\begin{smallmatrix}0\\-\sqrt{6}\end{smallmatrix}\bigl)$
                   &\ldots
\\\bottomrule[1pt]\bottomrule[1pt]
\end{tabular}}
\end{table}

\subsection{The M1 radiative decay behaviors and the magnetic moments}

{In this work, we further discuss the  M1 radiative decay behaviors and the magnetic moments of the possible $\Xi_{cc}K^{(*)}$ molecules. If the possible $\Xi_{cc}K$ and $\Xi_{cc}K^{*}$ molecules can coexist, then we can estimate the M1 radiative decay width for the $\Xi_{cc}K^*\to\Xi_{cc}K+\gamma$ process \cite{Wang:2022nqs}:
\begin{eqnarray}
\Gamma_{H\to H^{\prime}\gamma}&=&\frac{\alpha_{EM}}{2J_H+1}\frac{k^3}{m_p^2}\frac{\sum_{J_{H^\prime z},J_{Hz}}\left(\begin{array}{ccc}J_{H^\prime}&1&J_H\\-J_{H^\prime z}&0&J_{Hz}\end{array}\right)^2}{\left(\begin{array}{ccc}J_{H^\prime}&1&J_H\\-J_z&0&J_z\end{array}\right)^2}\nonumber\\
&&\times\frac{|\mu_{H\to H^\prime}|^2}{\mu_N^2}.\nonumber\\
\end{eqnarray}
In the above formula, $\alpha_{EM}\approx1/137$ is the electromagnetic fine structure constant, $k$ is the moment of the emitted photon with $k=(m_{H}^2-m_{H^{\prime}}^2)/2m_{H}$, $J_A$ and $J_{Az}$ with $A=(H, H')$ denoting the total and the $z-$component spin of the involved hadrons, respectively. $\mu_N=e/2m_p$ with $m_p = 0.938$ GeV \cite{ParticleDataGroup:2022pth}. $\left(\begin{array}{ccc}J_{H^\prime}&1&J_H\\-J_{H^\prime z}&0&J_{Hz}\end{array}\right)$ and $\left(\begin{array}{ccc}J_{H^\prime}&1&J_H\\-J_z&0&J_z\end{array}\right)$ stand for the $3-j$ coefficients, and $J_z$ is the minimum value of the total spin $J_A$ of the involved hadrons. $\mu_{H\to H^\prime}$ is the  corresponding transition magnetic moment, which can be expressed as
\begin{eqnarray}
\mu_{H\to H^\prime}&=&\left\langle J_{H^\prime},J_z\right|
     \sum_{j^\prime}\hat{\mu}_{zj^\prime}^{\text{spin}}e^{-i\bm{k}\cdot\bm{r}_{j^{\prime}}}
     +\hat{\mu}_{z}^{\text{orbital}}\left|J_H,J_z\right\rangle.
\end{eqnarray}
Here, $e^{-i\bm{k}\cdot\bm{r}_{j^{\prime}}}$ stands for the spatial wave function of the emitted photon. In our calculations, $e^{-i\bm{k}\cdot\bm{r}_{j^{\prime}}}$ is expanded as a serial of series containing the spherical Bessel functions $j_l(x)$ and the spherical harmonic functions $Y_{lm}(\Omega)$, i.e.,
\begin{eqnarray}
e^{-i\bm{k}\cdot\bm{r}_{j^{\prime}}} &=& \sum_{l=0}^{\infty}\sum_{m=-l}^{l}
    4\pi(-i)^lj_l(kr_{j^{\prime}})Y^*_{lm}(\Omega_{\bm{k}})Y_{lm}(\Omega_{{\bm{r}}_{j^{\prime}}}).
\end{eqnarray}
In order to gain a more comprehensive understanding of the spatial wave functions of the initial and final states, it is necessary to consider the spatial wave functions of the molecular states, the doubly charmed baryon, and the (anti-)kaon. In the case of the molecular states, we utilize the precise spatial wave functions by solving the Schr\"odinger equation. In the case of the doubly charmed baryon and the (anti-)kaon, we adopt the simple harmonic oscillator wave function, given by the expression $\psi_{n,l,m}(\beta,{\bf r})$, to describe their spatial wave functions. This can be written as
\begin{eqnarray}
\psi_{n,l,m}(\beta,{\bf r})&=&\sqrt{\frac{2n!}{\Gamma(n+l+\frac{3}{2})}}L_{n}^{l+\frac{1}{2}}(\beta^2r^2)\beta^{l+\frac{3}{2}}\nonumber\\
&&\times {\mathrm e}^{-\frac{\beta^2r^2}{2}}r^l Y_{l m}(\Omega).
\end{eqnarray}
In this context, the radial, orbital, and magnetic quantum numbers of the hadrons are designated as $n$, $l$, and $m$, respectively. $L_{n}^{l+\frac{1}{2}}(x)$ is used to represent the associated Laguerre polynomial. The oscillating parameters $\beta$ are taken to be $0.4~{\rm GeV}$ in the realistic calculations \cite{Wang:2022nqs}.

In the constituent quark model, the spin magnetic moment for all the involved quarks $(\hat{\mu}_z^{\text{spin}})$ and the orbital magnetic moment linking the molecular components $(\hat{\mu}_{z}^{\text{orbital}})$ can be expressed as follows \cite{Wang:2024sbw,Wang:2023ael,Lai:2024jfe,Wang:2022nqs}:
\begin{eqnarray}
\hat{\mu}_{j^{\prime}z}^{\text{spin}}&=&\frac{e_{j^{\prime}}\hat{\sigma}_{zj^{\prime}}}{2m_{j^{\prime}}},\\
\hat{\mu}_{z}^{\text{orbital}}&=&\left(\frac{m_M}{m_{B}+m_{M}}\frac{e_{B}}{2m_{B}}
    +\frac{m_B}{m_{B}+m_{M}}\frac{e_{M}}{2m_{M}}\right)\hat{L}_z.
\end{eqnarray}
Here, $e_{j^{\prime}, B, M}$ and $m_{j^{\prime}, B, M}$ are the charges and the masses of the $j^{\prime}$th constituent quarks of the hadron, the constituent baryon $B$ and meson $M$, respectively. $\hat{\sigma}_{zj^{\prime}}$ and $\hat{L}_z$ are the $z$-component of the Pauli spin operator of the $j^{\prime}$th constituent quarks of the hadron and the orbital angular momentum operator between the baryon and meson, respectively.

In general, the hadronic magnetic moment $\mu_H$ can be estimated by the expectation values of the $z-$component of the total magnetic moment operator $\hat{\mu}_z$, i.e.,
$\hat{\mu}_z=\sum_{j^{\prime}}\hat{\mu}_{j^{\prime}z}^{\text{spin}}+\hat{\mu}_{z}^{\text{orbital}}$. For the possible $\Xi_{cc}K^{(*)}$ molecules, their magnetic moments read as $\mu_H=\langle\Psi_H|\hat{\mu}_z|\Psi_H\rangle$, where $|\Psi_H\rangle$ is the wave function of the hadron $H$. For example, the magnetic moment for the $S-$wave $\Xi_{cc}K^{*}$ state with $I(J^P)=0(1/2^-)$ can be written as
\begin{eqnarray}
\mu_{\Xi_{cc}K^*}&=& \langle\Psi_{\Xi_{cc}K^*}|\hat{\mu}_z|\Psi_{\Xi_{cc}K^*}\rangle\nonumber\\
&=& \langle I',I_3'||I,I_3\rangle\langle\psi'(r)||\psi(r)\rangle\langle{}^2S_{1/2}|\hat{\mu}_z
|{}^2S_{1/2}\rangle.
\end{eqnarray}
Once we construct the spin and flavor wave functions for the doubly charmed baryon $\Xi_{cc}[ccq]$ and strange mesons $K^{(*)}[q\bar s]$, one can easily write down the corresponding magnetic moments. As collected in Table \ref{tab03}, we summarize the spin and flavor wave functions and the magnetic moments for the doubly charmed baryon $\Xi_{cc}$ and strange mesons $K^{(*)}$. The masses of the constituent quarks can be taken as $m_u=0.336$ GeV, $m_d=0.336$ GeV, $m_s=0.450$ GeV, and $m_c=1.680$ GeV \cite{Kumar:2005ei}.

\begin{table}[!htbb]
\renewcommand\tabcolsep{0.2cm}
\renewcommand{\arraystretch}{1.0}
\caption{The spin and flavor wave functions and the magnetic moments for the doubly charmed baryon $\Xi_{cc}$ and strange mesons $K^{(*)}$.}\label{tab03}
\begin{tabular}{lccc}
\toprule[1pt]\toprule[1pt]
Hadron &$\left|S,S_z\right\rangle$ &$\left|I,I_3\right\rangle$ &$\mu_H$\\\hline
{$\Xi_{cc}^{+}$}  &$\left|1/2,1/2\right\rangle$ &{$\left|1/2,-1/2\right\rangle$} &$\frac{4}{3}\mu_c-\frac{1}{3}\mu_d$ \\
                  &$\left|1/2,-1/2\right\rangle$ & &$-\frac{4}{3}\mu_c+\frac{1}{3}\mu_d$ \\\\
{$\Xi_{cc}^{++}$} &$\left|1/2,1/2\right\rangle$ &{$\left|1/2,1/2\right\rangle$} &$\frac{4}{3}\mu_c-\frac{1}{3}\mu_u$ \\
                  &$\left|1/2,-1/2\right\rangle$ & &$-\frac{4}{3}\mu_c+\frac{1}{3}\mu_u$ \\\\
{$K^{*0}$} &$\left|1,1\right\rangle$ &{$\left|1/2,-1/2\right\rangle$} &$\mu_d+\mu_{\bar{s}}$ \\
                  &$\left|1,0\right\rangle$ & &$0$ \\
                  &$\left|1,-1\right\rangle$ & &$-\mu_d-\mu_{\bar{s}}$ \\\\
{$K^{*+}$} &$\left|1,1\right\rangle$ &{$\left|1/2,1/2\right\rangle$} &$\mu_{\bar{s}}+\mu_{u}$ \\
                  &$\left|1,0\right\rangle$& &$0$  \\
                  &$\left|1,-1\right\rangle$ & &$-\mu_{\bar{s}}-\mu_{u}$\\\\
$K^0/K^+$         &$\left|0,0\right\rangle$ &$\left|1/2,-1/2(1/2)\right\rangle$  &$0$
\\\bottomrule[1pt]\bottomrule[1pt]
\end{tabular}
\end{table}}

\section{Numerical results}\label{sec3}

After deriving the total OBE effective potentials, we search for the bound state solutions (binding energy $E$ and root-mean-square (RMS) radius $r_{RMS}$) for the $\Xi_{cc}K^{(*)}$ and $\Xi_{cc}\bar K^{(*)}$ systems by solving the coupled channel Schr\"{o}dinger equations. And we vary the cutoff in the range from 0.80 GeV to 2.00 GeV. For a hadronic molecular state, reasonable bound state solutions have a binding energy $E$ ranging from several to several tens of MeV and an RMS radius $r_{RMS}$ around 1.00 fm or larger.

\subsection{The $\Xi_{cc}K^{(*)}$ systems}

In this subsection, we first study the mass spectrum behavior of the $\Xi_{cc}K^{(*)}$. We present our numerical results in six cases:
\begin{enumerate}
  \item[(i)]  cases I, II, and III correspond to the numerical results after considering the one-pion-exchange (OPE) interactions, the scalar together with vector meson exchange (OSVE) interactions, and the OBE interactions in the single-channel analysis, respectively.
  \item[(ii)]  cases IV, V, and VI correspond to the numerical results after considering the OPE, the OSVE, and the OBE interactions in the coupled-channel analysis, respectively.
\end{enumerate}
With these discussions, one can further explore the roles of the OPE, the OSVE interactions, and the coupled-channel effects in forming the $\Xi_{cc}K^{(*)}$ bound states.

For the $\Xi_{cc}K$ systems, the vertex ${K}-{K}-\pi$ violates the spin-parity conservation, so there do not exist the OPE interactions as shown in Eq. (\ref{potential1}). Therefore, one cannot obtain the bound state solutions in case I. The bound state solutions for the single $\Xi_{cc}K$ system in case II are the same as those in case III.

When we take the cutoff value from 0.80 GeV to 2.00 GeV, we cannot find the loosely bound state solutions for the single $\Xi_{cc}K$ states. Therefore, the OSVE interactions cannot provide strong enough interactions to bind the single $\Xi_{cc}K$ systems.

When we take into account the coupled channel effects, there are three channels, the $\Xi_{cc}K({}^2S_{1/2})$, $\Xi_{cc}K^*({}^2S_{1/2})$, and $\Xi_{cc}K^*({}^4D_{1/2})$ channels. The OPE interactions exist in the $\Xi_{cc}K\to\Xi_{cc}K^*$ and the $\Xi_{cc}K^*\to\Xi_{cc}K^*$ processes. As shown in Table \ref{num1}, we collect the bound state solutions for the coupled $\Xi_{cc}K/\Xi_{cc}K^*$ systems with $I(J^P)=0(1/2^-)$ in cases IV, V, and VI. Here, we find that for the coupled $\Xi_{cc}K/\Xi_{cc}K^*$ system with $I(J^P)=0(1/2^-)$, the loosely bound state solutions in cases V and VI emerge at the cutoff value around 1.70 GeV, and 1.45 GeV, respectively. The dominant channel is the $\Xi_{cc}K({}^2S_{1/2})$ channel. In case IV, there do not exist loosely bound state solutions as cutoff $\Lambda$ is less than 2.00 GeV. Therefore, the solo OPE interactions cannot provide the strong enough interactions to bind the coupled $\Xi_{cc}K/\Xi_{cc}K^*$ system with $I(J^P)=0(1/2^-)$. Compared to the OPE interactions, the OSVE interactions are much stronger attractive. When we consider the OBE interactions, the coupled $\Xi_{cc}K/\Xi_{cc}K^*$ system with $I(J^P)=0(1/2^-)$ can be recommended as a good molecular candidate, as the cutoff value is very close to the reasonable range $\Lambda\sim1.00$ GeV. In addition, the coupled channel effects play an important role in bind this coupled channel bound state.

In addition, for the coupled $\Xi_{cc}K/\Xi_{cc}K^*$ system with $I(J^P)=1(1/2^-)$, we cannot find the bound state solutions in cases IV, V, and VI.

\begin{table*}[htbp]
\renewcommand\tabcolsep{0.4cm}
\renewcommand{\arraystretch}{1.1}
\caption{The bound state properties (the binding energy $E$, the mass respected to the lowest channels $M$, the rms radius $r_{RMS}$, the probabilities for all the discussed channels) for the coupled $\Xi_{cc}K/\Xi_{cc}K^*$ systems with $I(J^P)=0(1/2^-)$ and the single $\Xi_{cc}K^*$ systems with $0(1/2^-,3/2^-)$ after adopting the OPE, OSVE, and OBE effective potentials. Here, $E$, $r_{RMS}$, $M$ and $\Lambda$ are in units of MeV, fm, and GeV, respectively.} \label{num1}
\begin{tabular}{lccccccc}
\toprule[1pt]\toprule[1pt]
$I(J^{P})=0(1/2^{-})$             &$\Lambda$    &$E$       &$r_{RMS}$    &M              &$\Xi_{cc}K({}^2S_{1/2})$    &$\Xi_{cc}K^*({}^2S_{1/2})$     &$\Xi_{cc}K^*({}^4D_{1/2})$        \\\hline
Case I            &$1.96$       &$-0.29$    &$5.11$       &$4514.72$            &\ldots&$99.72$&$0.28$             \\
$$                &$1.98$       &$-0.66$    &$4.16$       &$4514.35$            &\ldots&$99.34$&$0.66$             \\
$$                &$2.00$       &$-1.15$    &$3.37$       &$4513.86$            &\ldots&$99.14$&$0.86$             \\\\
Case II           &$1.46$       &$-0.52$    &$4.65$       &$4514.49$            &\ldots&$99.99$&$0.01$             \\
$$                &$1.56$       &$-5.70$    &$1.74$       &$4509.31$            &\ldots&$99.96$&$0.04$             \\
$$                &$1.66$       &$-16.99$   &$1.07$       &$4498.02$            &\ldots&$99.92$&$0.08$             \\\\
Case III          &$1.14$       &$-0.30$    &$5.19$       &$4514.71$            &\ldots&$99.89$&$0.11$             \\
$$                &$1.21$       &$-4.83$    &$1.86$       &$4510.18$            &\ldots&$99.79$&$0.21$             \\
$$                &$1.28$       &$-14.80$   &$1.14$       &$4500.21$            &\ldots&$99.78$&$0.22$             \\\\
Case V            &$1.70$       &$-0.46$    &$5.24$       &$4116.59$            &$98.33$&$1.62$&$0.06$             \\
$$                &$1.73$       &$-3.86$    &$2.51$       &$4113.19$            &$94.66$&$5.20$&$0.14$             \\
$$                &$1.76$       &$-11.60$   &$1.47$       &$4105.45$            &$88.99$&$10.79$&$0.22$             \\\\
Case VI           &$1.43$       &$-0.54$    &$5.03$       &$4116.51$            &$96.46$&$3.53$&$0.01$             \\
$$                &$1.45$       &$-4.56$    &$2.26$       &$4112.49$            &$89.46$&$10.53$&$0.02$             \\
$$                &$1.47$       &$-13.01$   &$1.33$       &$4104.04$            &$81.00$&$18.98$&$0.02$             \\\\
$I(J^{P})=0(3/2^{-})$             &$\Lambda$    &$E$       &$r_{RMS}$    &M              &$\Xi_{cc}K^*({}^4S_{3/2})$    &$\Xi_{cc}K^*({}^2D_{3/2})$     &$\Xi_{cc}K^*({}^4D_{3/2})$        \\
Case II           &$1.90$       &$-0.32$    &$5.23$       &$4514.69$            &$99.94$&$0.01$&$0.05$             \\
$$                &$1.95$       &$-0.64$    &$4.43$       &$4514.37$            &$99.91$&$0.02$&$0.07$             \\
$$                &$2.00$       &$-1.03$    &$3.77$       &$4513.98$            &$99.88$&$0.03$&$0.09$             \\\hline
\bottomrule[1pt]\bottomrule[1pt]
\end{tabular}
\end{table*}

For the $\Xi_{cc}K^*$ systems, the discussed quantum number configurations include $I(J^P)=0(1/2^-)$, $1(1/2^-)$, $0(3/2^-)$, and $1(3/2^-)$. By varying the cutoff value in the range of $0.80\leq\Lambda\leq2.00$ GeV, we cannot obtain the bound state solutions for the $\Xi_{cc}K^*$ system with $1(1/2^-)$. For the remaining three $\Xi_{cc}K^*$ systems, we present their bound state solutions in Table \ref{num1}. If taking the reasonable cutoff around 1.00 GeV, then the $\Xi_{cc}K^*$ state with $0(1/2^-)$ can be suggested as a good molecular candidate, with the dominant channel being the $\Xi_{cc}K^*({}^2S_{1/2})$. The OSVE interactions provide stronger interactions than the OPE interactions, as the cutoff in case II is much closer to the reasonable cutoff value.

For the isoscalar $\Xi_{cc}K^*$ system with $3/2^-$, the interactions arising from the OPE, OSVE, and OBE processes are stronger than those in the isovector systems. However, the cutoff in the isoscalar system is still a little far away from the value $\Lambda\sim1.00$ GeV. Therefore, the likelihood of the isoscalar $\Xi_{cc}K^*$ systems with $3/2^-$ being a hadronic molecular candidate is rather limited. Moreover, it is evident that the OSVE interactions yield stronger attractive forces compared to the OPE interactions, particularly when considering the cutoffs in cases I and II.

In summary, we can predict two promising candidates for hadronic molecules: the coupled $\Xi_{cc}K/\Xi_{cc}K^*$ state with $I(J^P)=0(1/2^-)$ and the $\Xi_{cc}K^*$ state with $0(1/2^-)$. The formation of these two bound states is notably influenced by the OSVE interactions, while the coupled channel effects significantly contribute to generate the coupled $\Xi_{cc}K/\Xi_{cc}K^*$ state with $I(J^P)=0(1/2^-)$.

{After that, we study the radiative decay width for the single $\Xi_{cc}K^*[0(1/2^-)]$ molecule decaying into the coupled $\Xi_{cc}K/\Xi_{cc}K^*[0(1/2^-)]$ molecule and the photon. As is well known, the radiative decay widths are not closely related to the radial wave functions but the spin-orbit wave functions. As shown in the Table \ref{num1}, the probabilities of the $D-$wave channels are very tiny, which cannot significantly influence the transition magnetic moments and the magnetic moments \cite{Wang:2022nqs}. In the following, we adopt their wave functions in the $S-$wave channels to calculate the corresponding electromagnetic properties. In Fig.\ref{fig3}, we present the radiative decay width of the $\Xi_{cc}K^*[0(1/2^-)]\to\Xi_{cc}K/\Xi_{cc}K^*[0(1/2^-)]+\gamma$ process, and we vary the binding energies for the single $\Xi_{cc}K^*[0(1/2^-)]$ molecule and the coupled $\Xi_{cc}K/\Xi_{cc}K^*[0(1/2^-)]$ molecule simultaneously. In the binding energy region of $0$ to $-12$ MeV, the obtained radiative decay width of the $\Xi_{cc}K^*[0(1/2^-)]\to\Xi_{cc}K/\Xi_{cc}K^*[0(1/2^-)]+\gamma$ process is around a few keV.

\begin{figure}[!htbp]
\center
\includegraphics[width=3.3in]{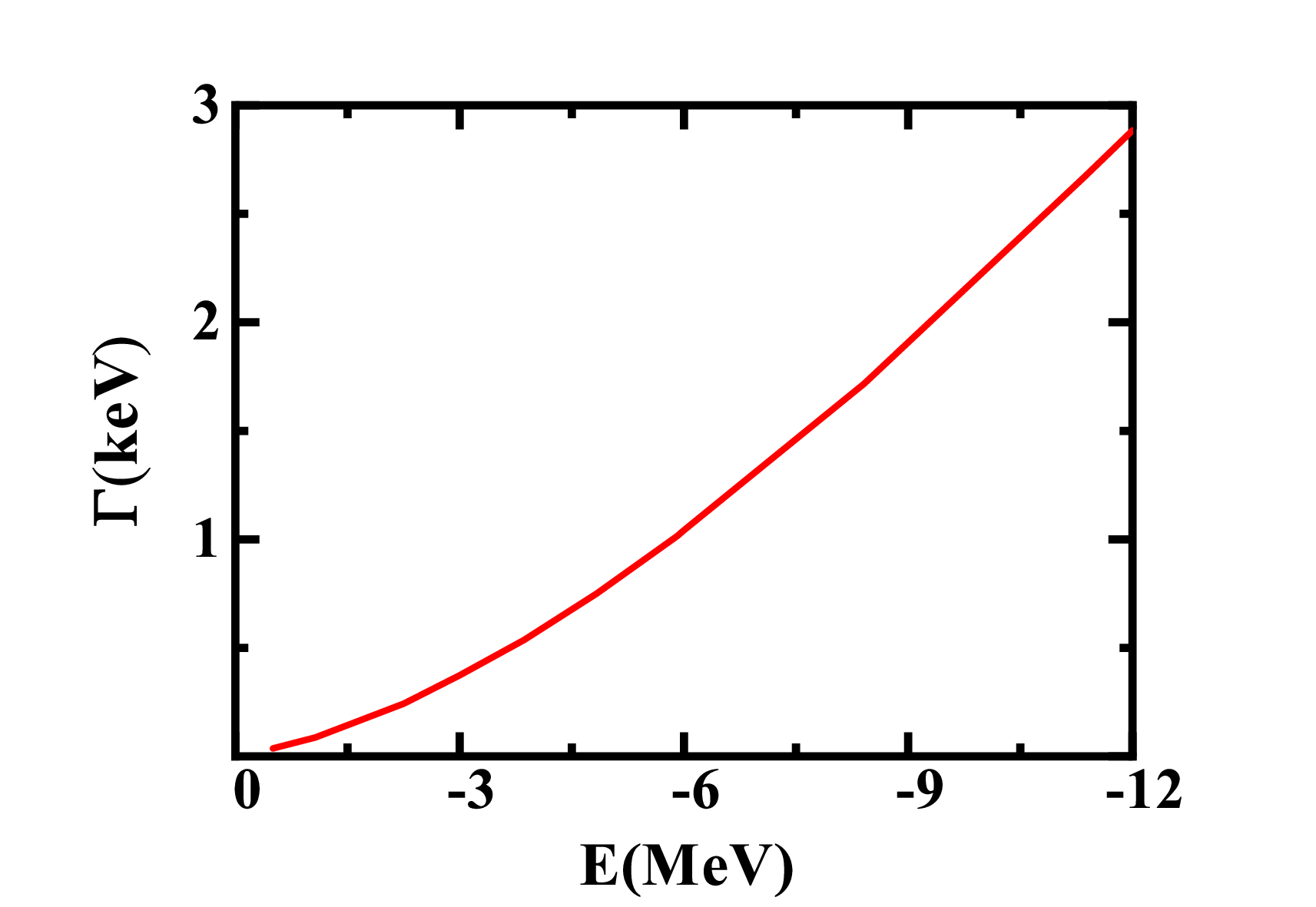}\\
\caption{The decay width of the $\Xi_{cc}K^*[0(1/2^-)]\to\Xi_{cc}K/\Xi_{cc}K^*[0(1/2^-)]+\gamma$ process.}\label{fig3}
\end{figure}

In addition, we further investigate the magnetic moments for the coupled $\Xi_{cc}K/\Xi_{cc}K^*$ molecule with $I(J^P)=0(1/2^-)$ and the $\Xi_{cc}K^*$ molecule with $0(1/2^-)$. As shown in Fig.\ref{fig2}, when we vary the binding energy from $0$ to $-12$ MeV, the magnetic moments for the coupled $\Xi_{cc}K/\Xi_{cc}K^*$ molecule with $I(J^P)=0(1/2^-)$ and the single $\Xi_{cc}K^*$ molecule with $0(1/2^-)$ are around 0.34 $\mu_N$ and 0.66 $\mu_N$, respectively. Our results indicate that the magnetic moments are insensitive with the corresponding binding energies.

\begin{figure}[!htbp]
\center
\includegraphics[width=3.3in]{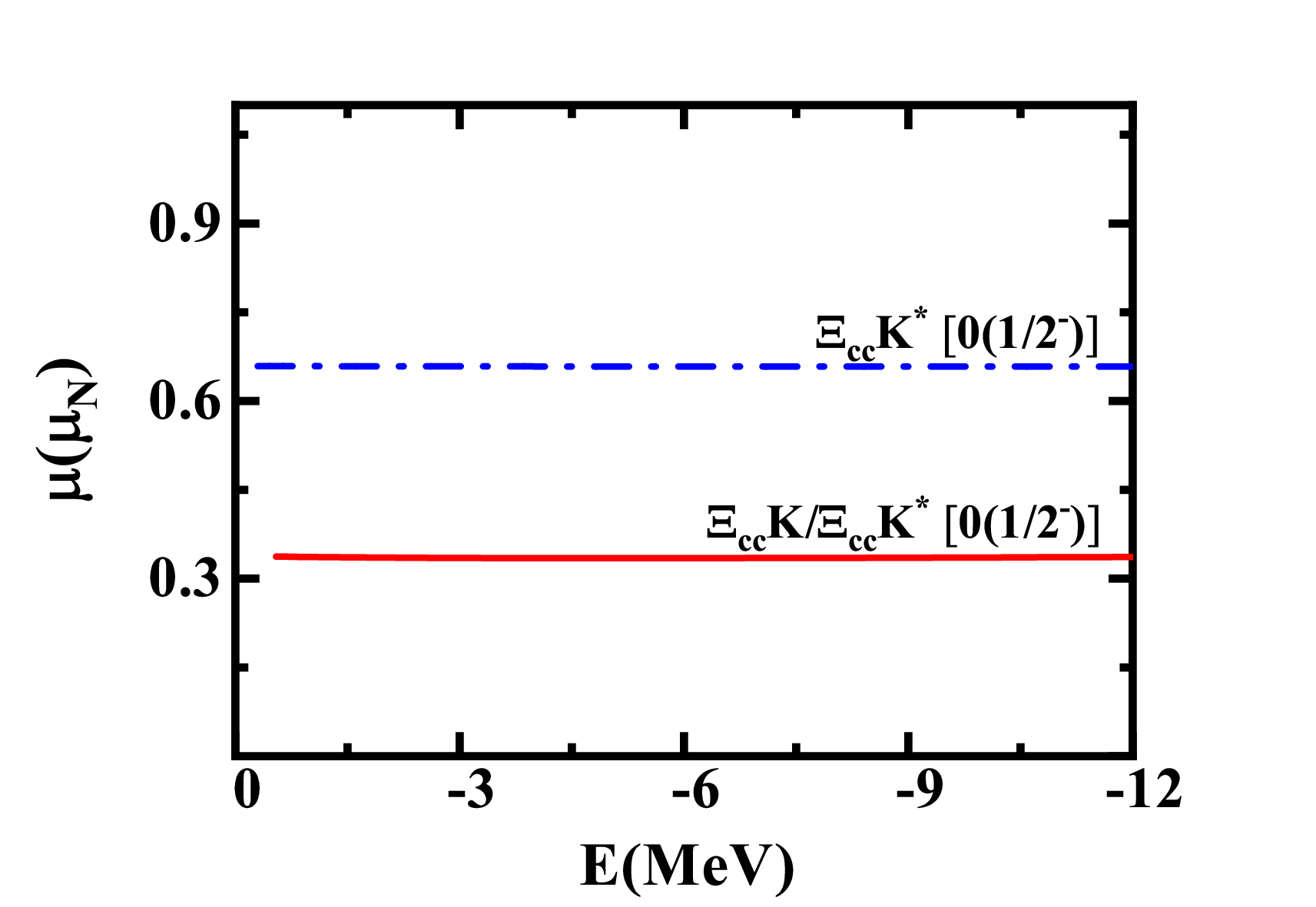}\\
\caption{The magnetic moments for the coupled $\Xi_{cc}K/\Xi_{cc}K^*$ molecule with $I(J^P)=0(1/2^-)$ and the $\Xi_{cc}K^*$ molecule with $0(1/2^-)$.}\label{fig2}
\end{figure}}

\subsection{The $\Xi_{cc}\bar{K}^{(*)}$ systems}

In this subsection, we extend our study to the $\Xi_{cc}\bar{K}^{(*)}$ interactions. The OBE effective potentials for the $\Xi_{cc}\bar{K}^{(*)}$ systems can be related to the $\Xi_{cc}K^{(*)}$ interactions by using the $G-$parity rule \cite{Chen:2021kad}, i.e., $V_E^{\Xi_{cc}\bar K^{(*)}\to\Xi_{cc}\bar K^{(*)}}=(-1)^{G_E}V_E^{\Xi_{cc}K^{(*)}\to\Xi_{cc}K^{(*)}}$, where $G_E$ is the $G-$parity for the exchanged mesons. Following the same procedures, we can first search for possible $\Xi_{cc}\bar{K}^{(*)}$ molecular candidates, and then explore their electromagnetic properties.

\begin{figure}[!htbp]
\center
\includegraphics[width=3.3in]{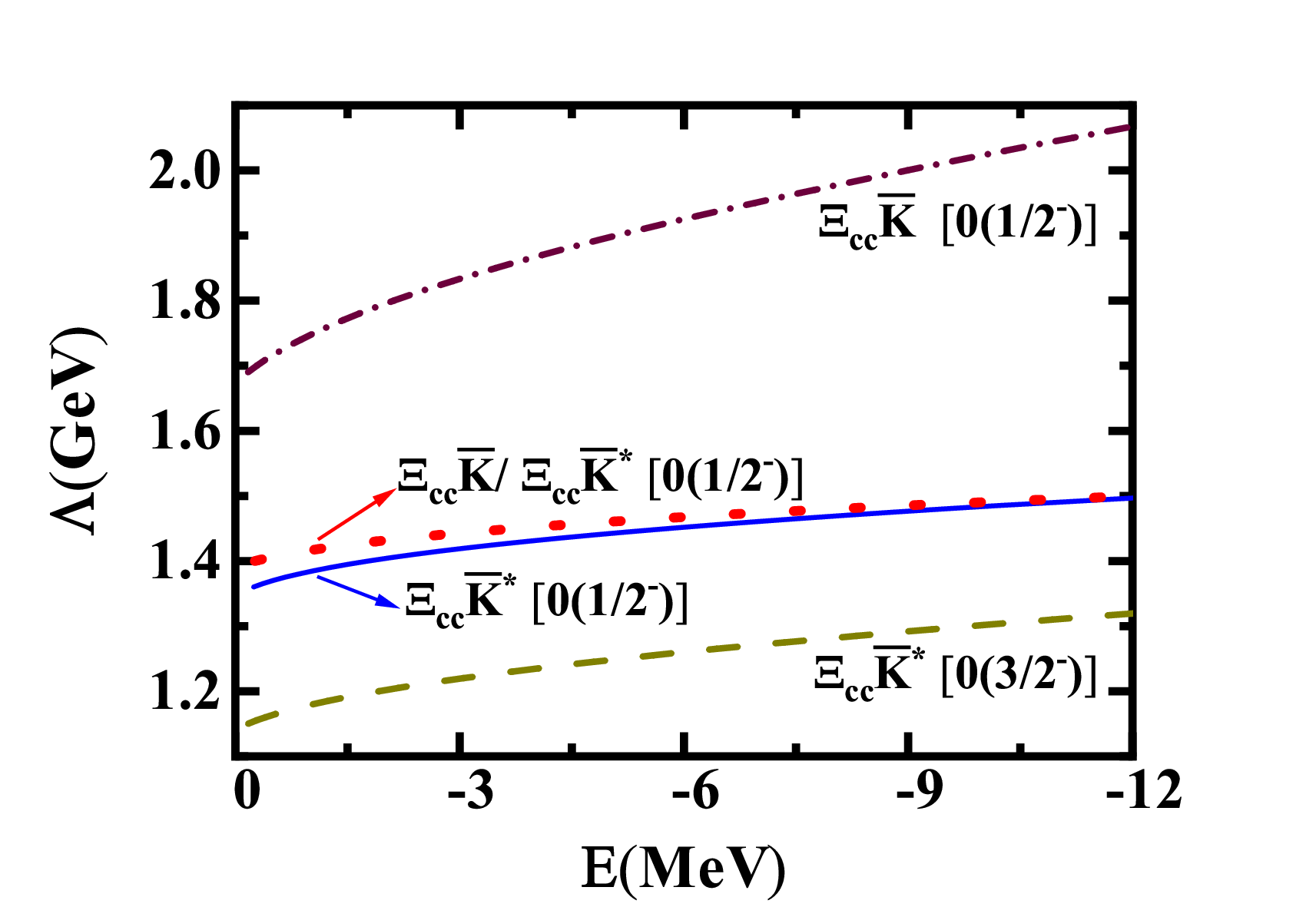}\\
\caption{The binding energies dependence of the cutoff for the single $\Xi_{cc}\bar K$ bound state with $0(1/2^-)$, the coupled $\Xi_{cc}\bar{K}/\Xi_{cc}\bar{K}^*$ bound state with $0(1/2^-)$, and the single $\Xi_{cc}\bar{K}^*$ bound states with $0(1/2^-)$ and $0(3/2^-)$.}\label{fig4}
\end{figure}

When we adopt the OBE effective potentials and vary the cutoff around 1.00 GeV, we can obtain four loosely bound states, the single $\Xi_{cc}\bar K$ bound state with $0(1/2^-)$, the coupled $\Xi_{cc}\bar{K}/\Xi_{cc}\bar{K}^*$ bound state with $0(1/2^-)$, and the single $\Xi_{cc}\bar{K}^*$ bound states with $0(1/2^-)$ and $0(3/2^-)$. In Fig.\ref{fig4}, we present the binding energies dependence of the cutoff values. It can be observed that the coupled channel effects play a positive role in generating the coupled $\Xi_{cc}\bar{K}/\Xi_{cc}\bar{K}^*$ bound state with $0(1/2^-)$. The OBE interactions for the $\Xi_{cc}K^*$ bound state with $0(1/2^-)$ are weaker attractive compared to the $\Xi_{cc}K^*$ interactions with $0(3/2^-)$ due to the larger cutoff values. The OBE interactions for the isovector $\Xi_{cc}\bar{K}^{(*)}$ systems cannot be strong enough to form loosely bound states with the reasonable cutoff values.

Based on the reasonable cutoff values, the coupled $\Xi_{cc}\bar{K}/\Xi_{cc}\bar{K}^*$ bound state with $0(1/2^-)$, and the single $\Xi_{cc}\bar{K}^*$ bound states with $0(1/2^-)$ and $0(3/2^-)$ can be recommended as the prime molecular candidates.

{In the following, we analyze the radiative decay widths and the magnetic moments of the coupled $\Xi_{cc}\bar{K}/\Xi_{cc}\bar{K}^*$ bound state with $0(1/2^-)$ and the single $\Xi_{cc}\bar{K}^*$ bound states with $0(1/2^-)$ and $0(3/2^-)$. As shown in Fig.\ref{fig6}, we find the radiative decay widths of the $\Xi_{cc}\bar K^*[0(1/2^-)]\to\Xi_{cc}\bar K/\Xi_{cc}\bar K^*[0(1/2^-)]+\gamma$ and $\Xi_{cc}\bar K^*[0(3/2^-)]\to\Xi_{cc}\bar K/\Xi_{cc}\bar K^*[0(1/2^-)]+\gamma$ processes are around several keV.

\begin{figure}[!htbp]
\center
\includegraphics[width=3.3in]{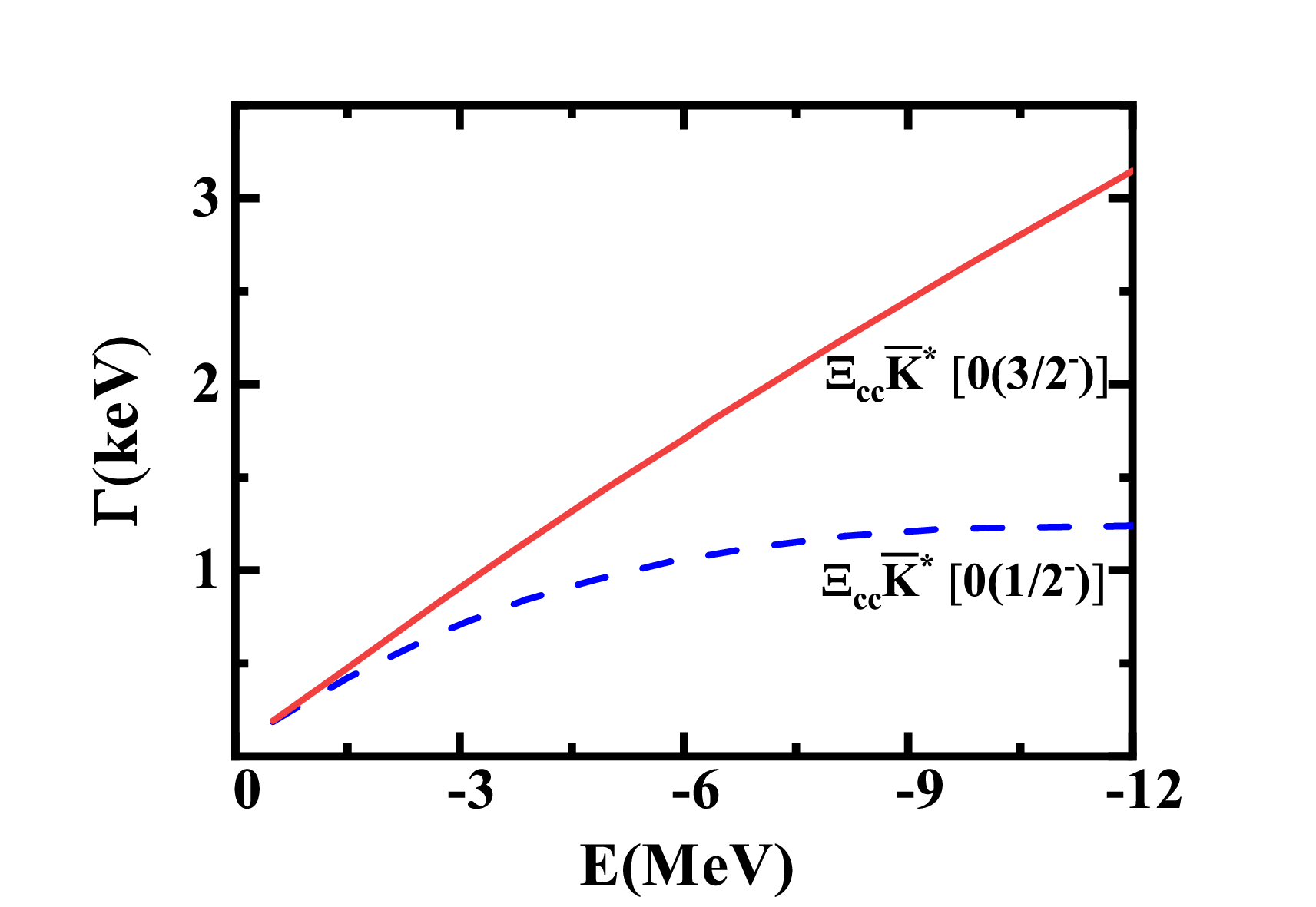}
\caption{The decay widths of the $\Xi_{cc}\bar K^*[0(1/2^-)]\to\Xi_{cc}\bar K/\Xi_{cc}\bar K^*[0(1/2^-)]+\gamma$ and $\Xi_{cc}\bar K^*[0(3/2^-)]\to\Xi_{cc}\bar K/\Xi_{cc}\bar K^*[0(1/2^-)]+\gamma$ processes.}\label{fig6}
\end{figure}

In Fig.\ref{fig5}, we present the binding energies dependence of the magnetic moments for the coupled $\Xi_{cc}\bar{K}/\Xi_{cc}\bar{K}^*$ bound state with $0(1/2^-)$ and the single $\Xi_{cc}\bar{K}^*$ bound states with $0(1/2^-)$ and $0(3/2^-)$. Here, we find the magnetic moments for the $\Xi_{cc}\bar{K}^*$ molecules with $0(1/2^-)$ and $0(3/2^-)$ are around $-0.87$ $\mu_N$ and $-0.81$ $\mu_N$, respectively. In addition, the magnetic moment for the coupled $\Xi_{cc} \bar K/\Xi_{cc} \bar K^*$ molecule with $0(1/2^-)$ is around 0.34 $\mu_N$.

\begin{figure}[!htbp]
\center
\includegraphics[width=3.3in]{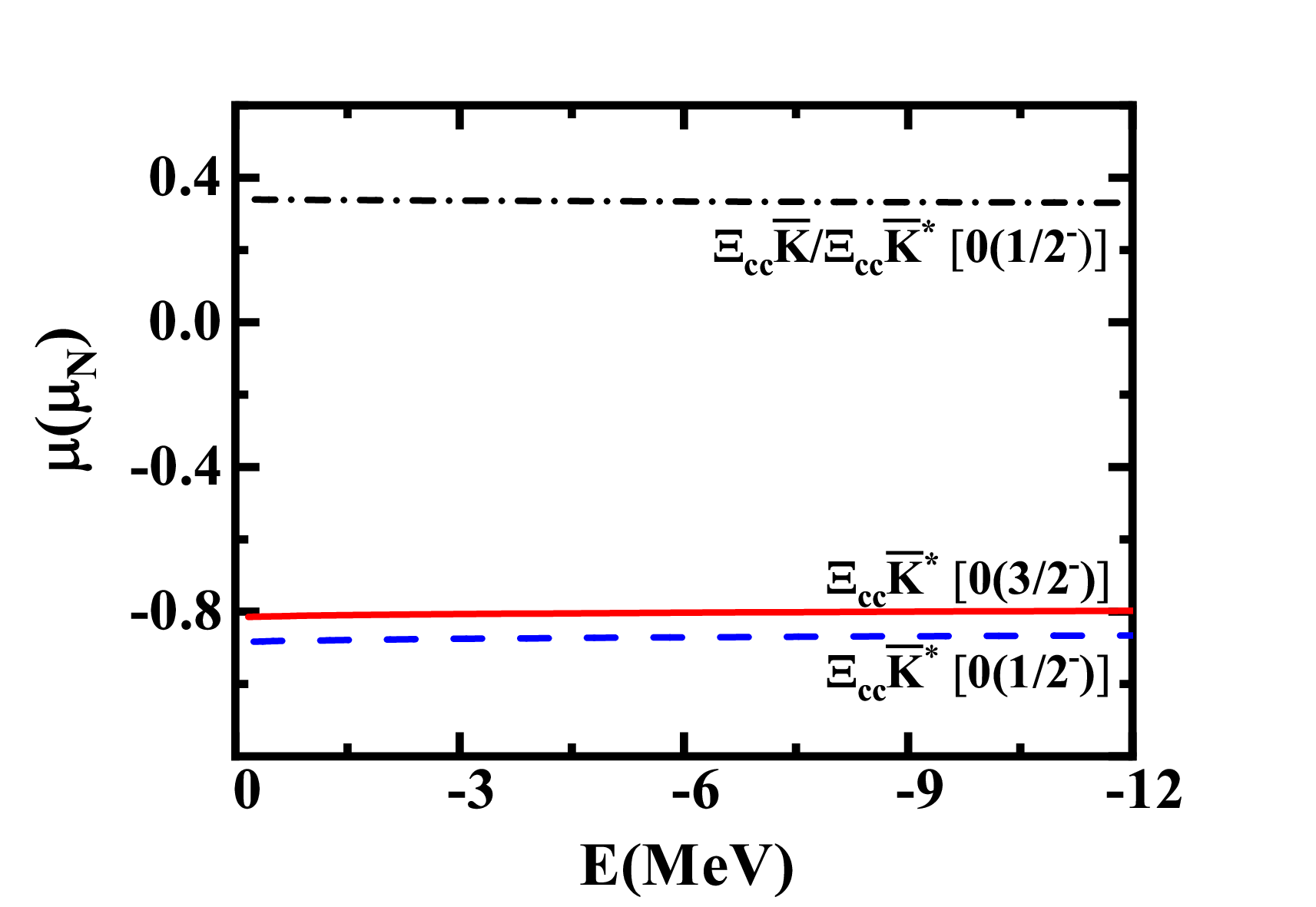}
\caption{The magnetic moments for the coupled $\Xi_{cc}\bar{K}/\Xi_{cc}\bar{K}^*$ bound state with $0(1/2^-)$, and the single $\Xi_{cc}\bar{K}^*$ bound states with $0(1/2^-)$ and $0(3/2^-)$.}\label{fig5}
\end{figure}}

\section{Summary}\label{sec4}

The study of hadron spectroscopy is a fundamental issue in nonperturbative strong interactions. Recent experimental observations, such as the doubly charmed structures $\Xi_{cc}^{++}(3621)$ \cite{LHCb:2017iph} and $T_{cc}$ \cite{LHCb:2021vvq,LHCb:2021auc}, have sparked significant interest in exploring other possible doubly charmed states. Theoretical investigations have proposed various molecular states and examined their formation mechanisms and properties \cite{Chen:2021vhg,Wang:2023ael,Feijoo:2021ppq,Wang:2021yld,Chen:2019asm,Wang:2023eng,Yan:2023iie,Wang:2023aob}. In this work, we focus on possible doubly charmed hadronic molecular states with strangeness $|S|=1$ primarily composed of the $S$-wave $\Xi_{cc}K^{(*)}$ and $\Xi_{cc}\bar{K}^{(*)}$ channels. We adopt the OBE model to study the relevant interactions. When deducing the OBE effective potentials, we consider both the $S-D$ wave mixing effects and the coupled channel effects.

Finally, we can predict five primary hadronic molecular candidates: the coupled $\Xi_{cc}K/\Xi_{cc}K^*$ molecule with $I(J^P)=0(1/2^-)$, the $\Xi_{cc}K^*$ molecule with $0(1/2^-)$, the $\Xi_{cc}\bar{K}$ molecule with $0(1/2^-)$, and the $\Xi_{cc}\bar{K}^*$ molecules with $0(1/2^-,3/2^-)$. The coupled channel effects play an essential role in generating the coupled $\Xi_{cc}K/\Xi_{cc}K^*$ molecule with $0(1/2^-)$.

Afterwards, we employ the constituent quark model to investigate the electromagnetic properties of these predicted molecular candidates, including their M1 radiative decay behaviors and magnetic moments between these molecular candidates. Our results indicate that the magnetic moments of molecular states are almost independent of their binding energies. The magnetic moments for the coupled $\Xi_{cc}K/\Xi_{cc}K^*$ molecule with $0(1/2^-)$ and the coupled $\Xi_{cc} \bar K/\Xi_{cc} \bar K^*$ molecule with $0(1/2^-)$ are around 0.34 $\mu_N$, stemming from the doubly charmed baryon. For the $\Xi_{cc}K^*$ molecule with $0(1/2^-)$, the $\Xi_{cc}\bar{K}^*$ molecules with $0(1/2^-)$ and $0(3/2^-)$, the magnetic moments are around 0.66 $\mu_N$, $-0.87$ $\mu_N$, and $-0.81$ $\mu_N$, respectively. The decay widths for the $\Xi_{cc}K^*[0(1/2^-)]\to\Xi_{cc}K/\Xi_{cc}K^*[0(1/2^-)]+\gamma$, $\Xi_{cc}\bar{K}^*[0(1/2^-,3/2^-)]\to\Xi_{cc}\bar K/\Xi_{cc}\bar K^*[0(1/2^-)]+\gamma$ processes are around a few keV.

Overall, our investigations can open up new avenues for exploring the existence of possible doubly charmed molecular pentaquarks with strangeness $|S|=1$. We look forward to experimental confirmation of these predictions.

\section*{ACKNOWLEDGMENTS}

R.C. is supported by the National Natural Science Foundation of China under Grant No. 12305139 and the Xiaoxiang Scholars Programme of Hunan Normal University. X.L. is supported by the National Natural Science Foundation of China under Grants No. 12335001 and No. 12247101, National Key Research and Development Program of China under Contract No. 2020YFA0406400, the 111 Project under Grant No. B20063, the fundamental Research Funds for the Central Universities, and the project for top-notch innovative talents of Gansu province. F.L.W. is supported by the China Postdoctoral Science Foundation under Grant No. 2022M721440.

\end{document}